\documentclass[aps,a4,twocolumn,showpacs]{revtex4-1}
\usepackage{amsmath}
\usepackage{latexsym}
\usepackage{float}
\usepackage{amssymb}
\usepackage{graphicx}
\usepackage{epsfig}

\begin{document}

\title{Polymer chain stiffness versus excluded volume: A Monte Carlo study of the
crossover towards the wormlike chain model}

\author{Hsiao-Ping Hsu$^a$, Wolfgang Paul$^b$, and Kurt Binder$^a$}

\affiliation{$^a$Institut f\"ur Physik, Johannes Gutenberg-Universit\"at Mainz,\\
 Staudinger Weg 7, D-55099 Mainz, Germany \\
$^b$Theoretische Physik, Martin Luther Universit\"at \\
Halle-Wittenberg, von Seckendorffplatz 1, 06120 Halle, Germany}


\begin{abstract}
When the local intrinsic stiffness of a polymer chain varies over
a wide range, one can observe both a crossover from rigid-rod-like
behavior to (almost) Gaussian random coils and a further crossover
towards self-avoiding walks in good solvents. Using the
pruned-enriched Rosenbluth method (PERM) to study self-avoiding
walks of up to $N_b=50000$ steps and variable flexibility, the
applicability of the Kratky-Porod model is tested. Evidence for
non-exponential decay of the bond-orientational correlations
$\langle \cos \theta (s) \rangle$ for large distances $s$ along
the chain contour is presented, irrespective of chain stiffness.
For bottle-brush polymers on the other hand, where experimentally stiffness
is varied via the length of side-chains, it is shown that
these cylindrical brushes (with flexible backbones)
are not described by the Kratky-Porod wormlike chain
model, since their persistence length is (roughly) proportional to
their cross-sectional radius, for all conditions of practical
interest.
\end{abstract}

\maketitle

\section{Introduction and motivation}
One of the most basic characteristics of macromolecules with
linear ``chemical architecture'' is chain flexibility (or lack
thereof, stiffness) \cite{1,2,3}. While many synthetic polymers
are fully flexible, and the statistical properties of their
conformations under good solvent conditions have been extensively
investigated \cite{1,2,3,4,5}, recently also semiflexible polymers
have found much interest, in particular since important
biopolymers such as DNA, some proteins, rodlike viruses, or actin
filaments belong to this class \cite{6,7,8}. Moreover, also
synthetic polymers such as polyethylene exhibit some stiffness
over short distances along the chain. Particular interest in the
problem of chain stiffness has arisen due to the discovery that in
macromolecules with ``bottle-brush architecture'' \cite{9,10,11,12}
their stiffness can be varied over a wide range by changing the
length and grafting density of side chains. Since short
bottle-brush polymers in solution may exhibit liquid crystalline
type ordering \cite{13}, and their structure is very sensitive to
various external stimuli, various applications for such molecules
have been proposed (from building blocks of nanostructures to
actuators and sensors etc. \cite{14,15}). Also in a biological
context semiflexible macromolecules with bottle-brush architecture
are found \cite{16}, and have interesting functions such as
lubrication in mammalian joints \cite{17}.

While for fully flexible polymers under good solvent conditions
excluded volume interactions (loosely speaking, monomers of a
polymer chain cannot ``sit'' on top of each other) are of central
importance \cite{1,2,3,4,5}, the standard model to describe
semiflexible chains ignores them completely \cite{18,19,20,21,22}.
This ``standard model'' is the ``wormlike chain model'' of Kratky
and Porod \cite{18}, described by a Hamiltonian (in the continuum
limit)

\begin{equation} \label{eq1}
\mathcal{H}= \frac{\kappa}{2} \int\limits_0^L dt \Big(\frac{d^2
\vec{r}(t)}{dt^2}\Big)^2 \; ,
\end{equation}
where the curve $\vec{r}(t)$ describes the linear polymer of
contour length $L$, and the parameter $\kappa$ describing the
bending stiffness is related to the ``persistence'' length $\ell_
p$ as $\kappa=\ell_p k_BT$. For chain molecules in the
absence of excluded volume, where distances between monomers that
are far apart satisfy Gaussian statistics, one can show that the
orientational correlation function between bond vectors decays
exponentially with $s=t/\ell_b$

\begin{equation} \label{eq2}
\langle \vec{a}_i \cdot \vec{a}_{i+s} \rangle = \ell_b^2 \exp (-s \ell_b/
\ell_p) \; ,
\end{equation}
where $\ell_b=|\vec{a}_i|$ is bond length between two subsequent
monomers along the chain, $\vec{a}_i=\vec{r}_i-\vec{r}_{i-1}$. 
Note that $s$ is dimensionless and for a model with discrete monomers
just denotes the difference in the labels $i$, $i+s$ of the monomers.
The Kratky-Porod model then yields for the end-to-end vector
$\vec{R}_e$ of the semiflexible chain \cite{18} $(L=N_b\ell_b)$

\begin{equation} \label{eq3}
\langle R^2 _e \rangle = 2 \ell _p L \Big\{1-\frac{\ell_p}{L} [1-\exp
(-L/\ell_p)]\Big \}
\end{equation}
which shows the standard Gaussian behavior $(\langle R^2_e \rangle
= 2\ell_p L= 2 \ell_p \ell_b N_b)$ for $L \rightarrow \infty$,
while for $L<<\ell_p$ the rod-like behavior $\langle R^2_e
\rangle=L^2$ results.

However, for good solvent conditions and long enough chains
eqs.~(\ref{eq2}), and (\ref{eq3}) cannot be correct: rather we must have
\cite{23}, for $N_b \rightarrow \infty$,

\begin{equation} \label{eq4}
\langle \vec{a}_i \cdot \vec{a}_{i+s} \rangle \propto s^{- \beta}
\;, \quad \beta=2-2 \nu \approx 0.824 \,\, , \, \, s^* \ll s
\ll N_b
\end{equation}
and \cite{1,2,3,4,5,24,25}

\begin{equation} \label{eq5}
\langle R^2_e \rangle= 2 \ell_{p,R} \ell_b N_b^{2 \nu}
\end{equation}
where $\nu \approx 0.588$ \cite{24} is the exponent describing the
swelling of the chains due to excluded volume. 
If $s$ is no longer much smaller than $N_b$, the power law
eq.~(\ref{eq4}), gradually crosses over to a faster decay
that depends on $N_b$~\cite{25}.
In eq.~(\ref{eq5}), 
we have written a tentative generalization of the Kratky-Porod
result to introduce another, effective persistence length
$\ell_{p,R}$ \cite{25}. The question that we ask in this paper
hence is, how can one reconcile eq.~(\ref{eq2}) with
eq.~(\ref{eq4}), as well as eq.~(\ref{eq3}) with eq.~(\ref{eq5})?
Tentatively, one might expect that for semiflexible chains
eq.~(\ref{eq2}) still holds up to some characteristic, large value
$s^*$, where then the crossover to eq.~(\ref{eq4}) takes place
\cite{26}; but such a hypothesis is by no means evident, and
remains to be proven, and the value of $s^*$ remains to be
predicted. Similarly, one might expect that eq.~(\ref{eq3}) holds
only for $N_b$ up to some value $N^*_b$, and then a crossover to
eq.~(\ref{eq5}) takes place. Using a Flory \cite{27} argument,
Netz and Andelman \cite{28} suggested that this is the case and
(implying $\nu=3/5$)

\begin{equation} \label{eq6}
N^*_b \propto(\ell_p/\ell_b)^3 \; , \quad \ell_{p,R} \propto
\ell^{2/5}_p \ell^{3/5}_b \; .
\end{equation}
As expected, this result for $\ell_{p,R}$ is equivalent to
the classical result~\cite{29a,30a,31a} 
$\langle R_e^2 \rangle^{1/2} \propto \ell_b (\ell_p/\ell_b)^{1/5}
N_b^{3/5}$ for semiflexible macromolecules in the limit
$N_b \rightarrow \infty$.

In the present work, we hence wish to check whether such ideas
about these crossovers apply, and if so, study their detailed
behavior. Finally, we shall discuss whether or not these ideas
have some bearing on the problem of the persistence length of
bottle-brush polymers \cite{25}.

\begin{figure}
\begin{center}
\vspace{1cm}
(a)\includegraphics[scale=0.25,angle=270]{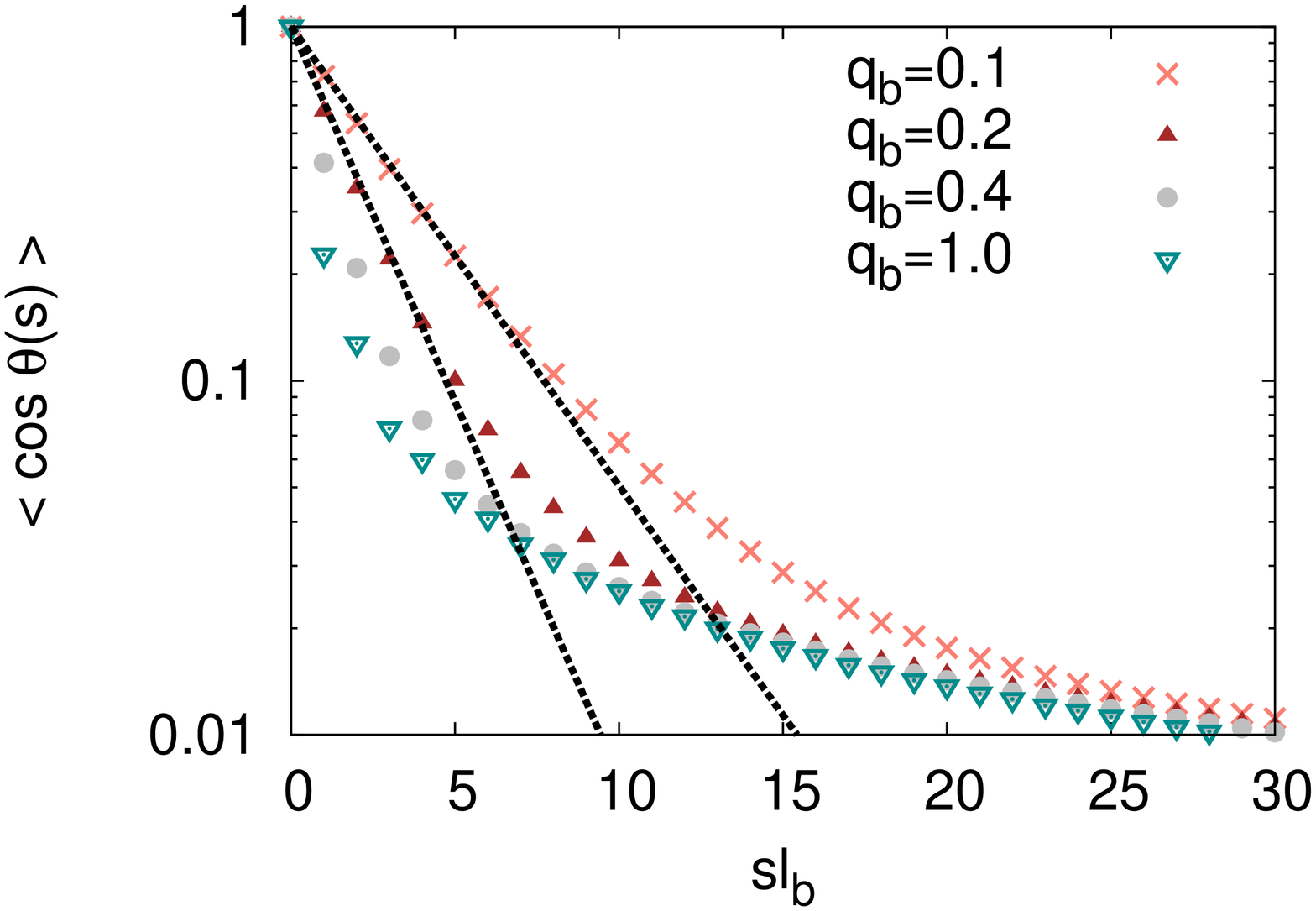}\\
(b)\includegraphics[scale=0.25,angle=270]{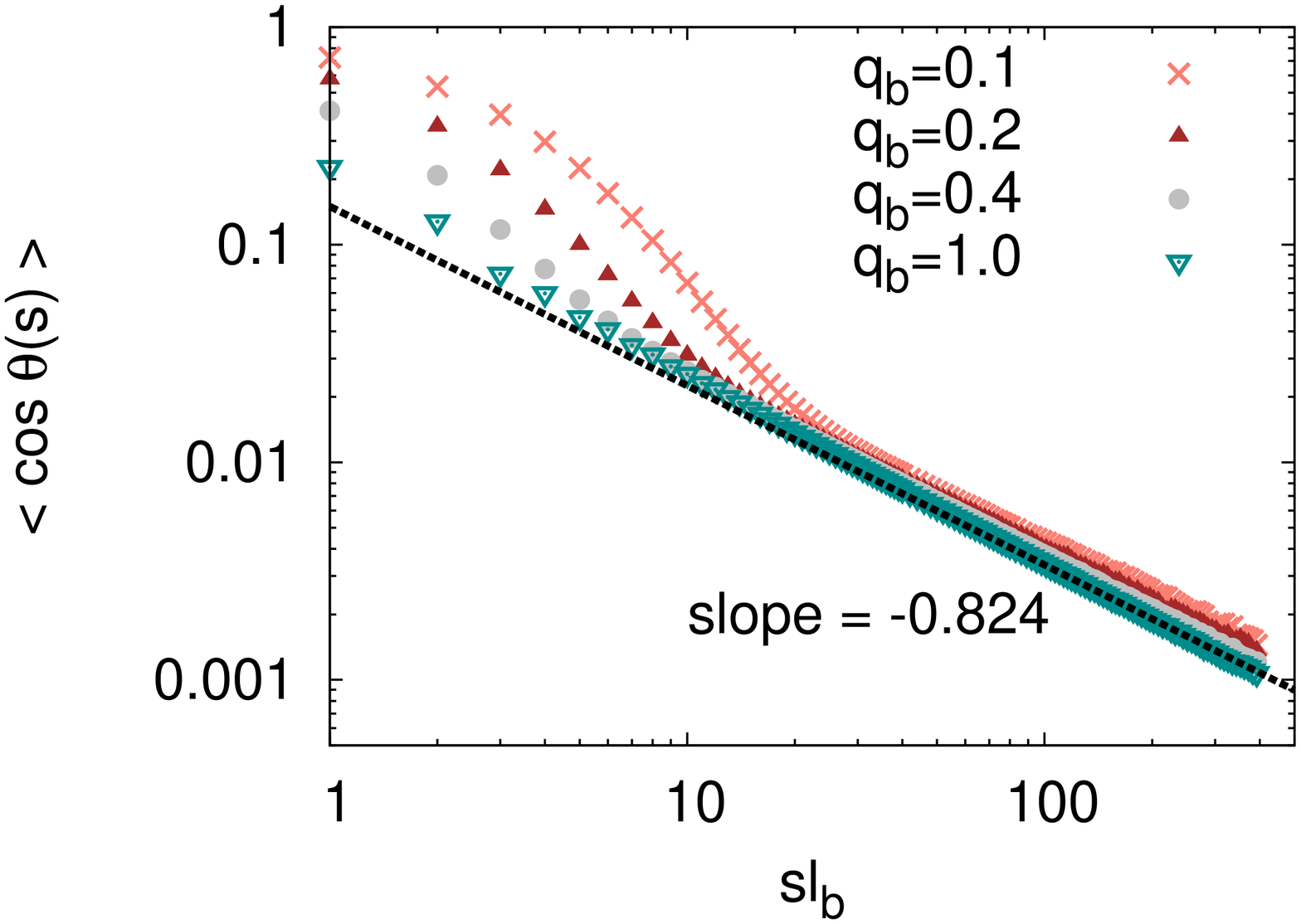}\\
(c)\includegraphics[scale=0.25,angle=270]{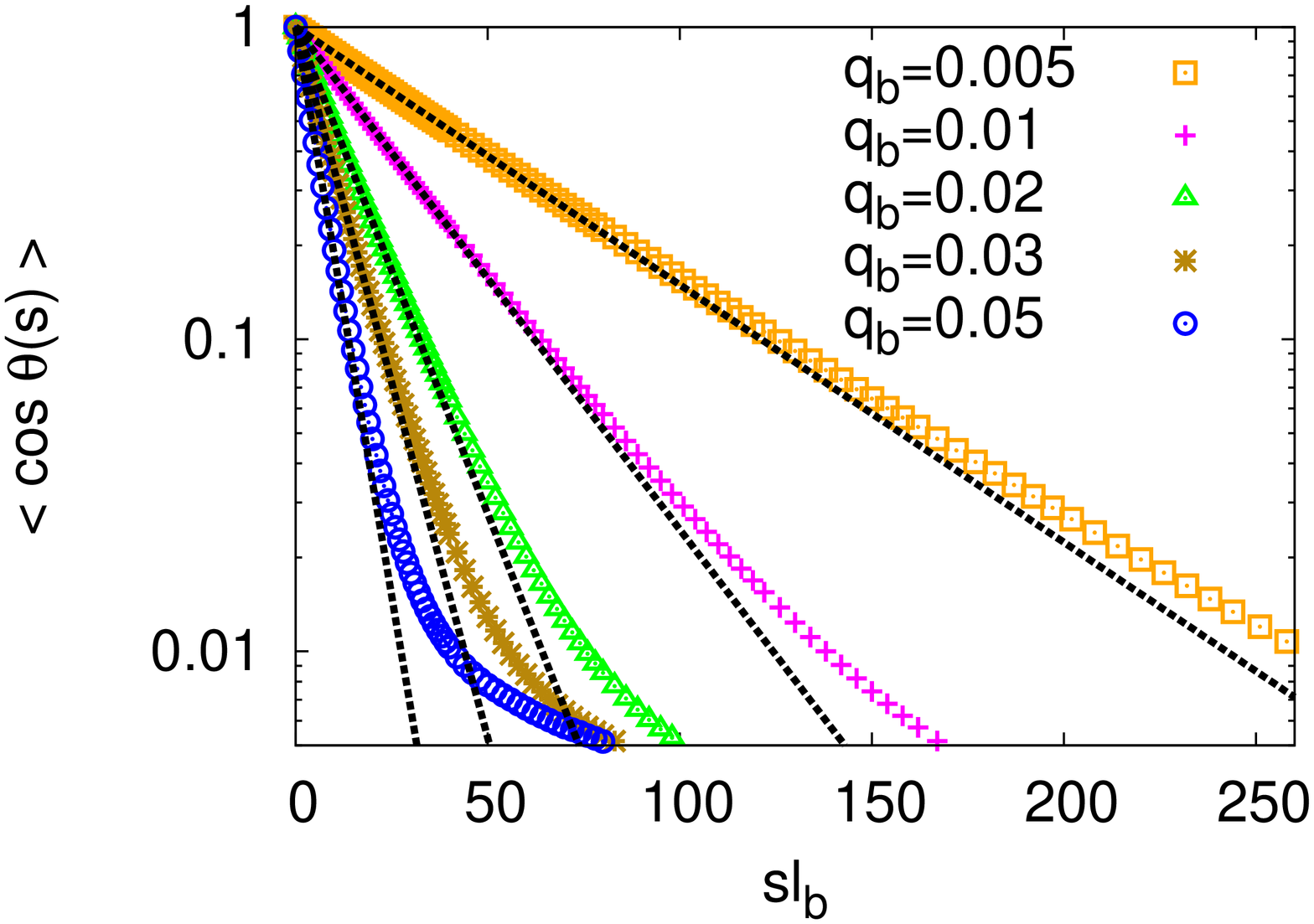}\\
(d)\includegraphics[scale=0.25,angle=270]{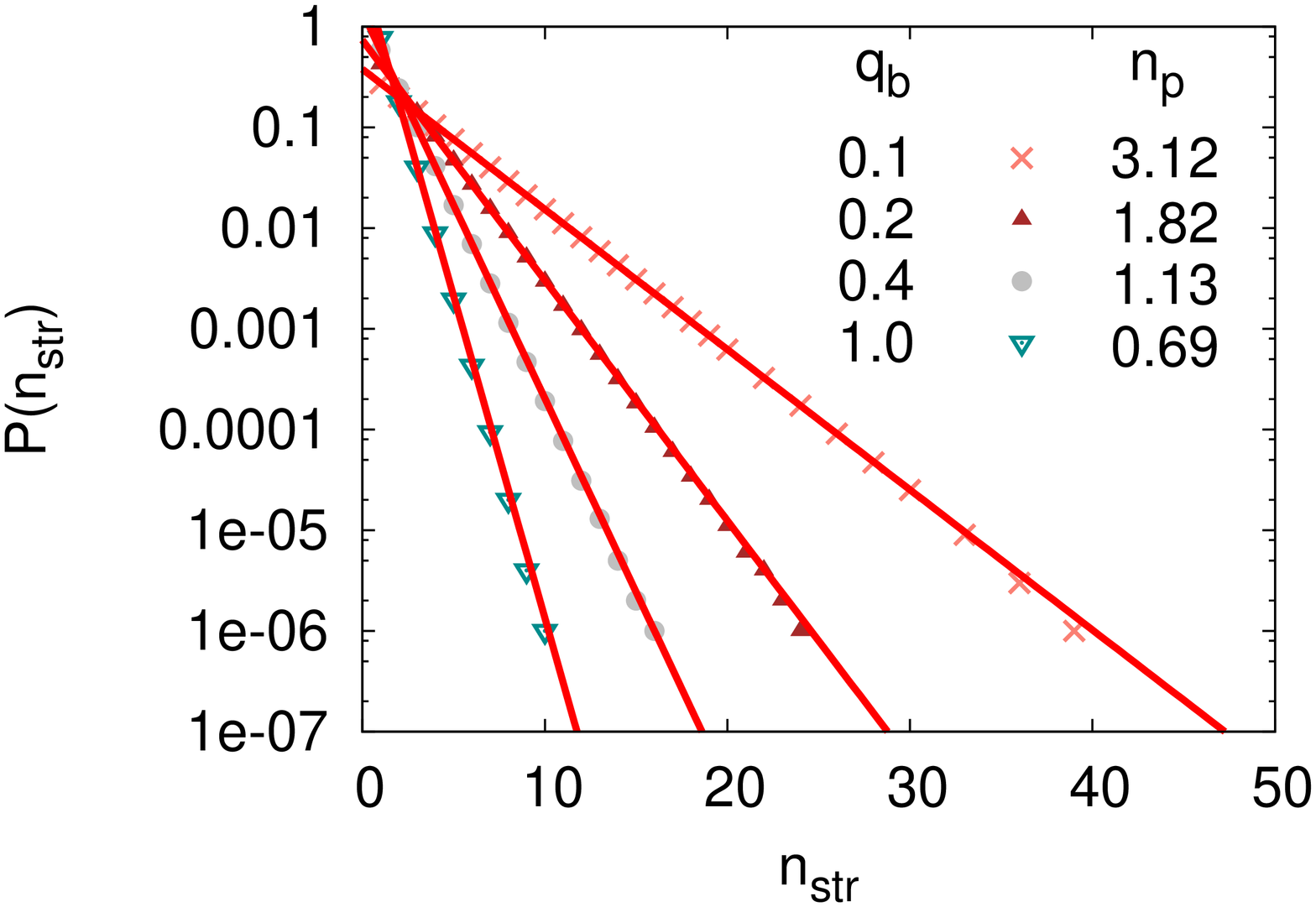}
\caption{Semi-log plot of the bond vector correlation function
$\langle \cos \theta (s) \rangle$ vs. the chemical distance
$\ell_bs$ along the chain (here $\ell_b$ is the lattice spacing
which is our unit of length,
$\ell_b=1$) for $0.1\leq q_b \leq 1.0$ (a), and for $0.005 \leq q_b
\leq 0.05 $ (c). Part (b) shows the same data as part (a) but on a
log-log plot. The straight line indicates a fit of eq.~(\ref{eq4})
to the data for $q_b=1$ while straight lines in (a) (c)
are fits of the initial exponential decay to eq.~(\ref{eq2}). 
Part (d) shows the distribution function
$P(n_{\rm str})$ of straight pieces of the chain without kinks,
together with fits of $P(n_{\rm str}) \propto \exp 
(-n_{\rm str}/n_p)$. $N_b=50000$ was used throughout.}
\label{fig1}
\end{center}
\end{figure}

\section{Model and simulation results}

We carried out Monte Carlo simulations of self-avoiding walks
(SAWs) on the simple cubic lattice, applying an energy
$\varepsilon _b (1-\cos \theta)$ if a bond orientation differs by
an angle $\theta$ relative to the preceding bond (of course, on
our lattice only $\theta =0$ or $\theta = \pm \pi/2$ is possible).
Using the pruned-enriched Rosenbluth method (PERM) \cite{29,30},
the partition function of the SAWs of $N_b$ steps with $N_{\rm
bend}$ local bends (where $\theta = \pm \pi/2)$ is written as

\begin{equation} \label{eq7}
Z_{N_b, N_{\rm bend}} (q_b)=\sum\limits_{\rm config} C(N_b, N_{\rm
bend})q^{N_{\rm bend}}_b
\end{equation}
where $q_b=\exp [- \varepsilon_b/k_BT]$ is the appropriate
Boltzmann factor ($q_b=1$ for ordinary SAW's). We obtained data
for $C(N_b, N_{\rm bend})$ for $N_b$ up to $N_b=50000$, and vary
$q_b$ over a wide range as well, 0.005$\leq q_b \leq 1.0$.
In addition we have continued our simulations~\cite{25} of
bottle-brush polymers using the bond fluctuation model.

Fig.~\ref{fig1} shows that two distinct patterns of behavior
emerge: for flexible or only moderately stiff chains, $0.1 \leq
q_b \leq1.0$, eq.~(\ref{eq2}) has hardly any significant regime of
applicability, there are just a few discrete values for small
$s (s=1,2,3,\cdots)$ that can be fitted by eq.~(\ref{eq2}), and
then pronounced deviations from a simple exponential behavior
occur (fig.~\ref{fig1}a). On the other hand, for large $s$ 
$(s \geq 20)$ the power law \{eq.~(\ref{eq4})\} provides a good fit
(fig.~\ref{fig1}b). But it is also evident that the deviations
from the power law for small $s$ become the more pronounced the
smaller $q_b$ is. Of course, the crossover value $s^*$ is not
sharply defined, but rather the crossover is smeared out over some
range in $s$. For small $q_b$ $(0.005 \leq q_b \leq 0.05$) the
initial exponential decay \{eq.~(\ref{eq2})\} becomes better
visible, fig.~\ref{fig1}c, and chain lengths even larger than
studied by us would be needed to still clearly identify the power
law for large enough $s$. While the exponential decay of the bond
correlations \{eq.~(\ref{eq2})\} works only for a restricted range of
$s (s \ll s^*)$ and only for very stiff chains ($q_b \ll 1)$, a
simple exponential decay always is found for the probability
$P(n_{\rm str})$ that sequences of $n_{\rm str}$ subsequent bonds
without kink occur, for large $n_{\rm str}$ (fig.~\ref{fig1}d,
$P(n_{\rm str}) \propto \exp(-n_{\rm str}/n_p)$
where $n_p$ then is defined as decay constant. For $q_b \ll 1$ we
find that $\ell_p=n_p \ell_b$, within numerical error, and
furthermore $\ell_p \propto q^{-1}_b$ holds then.

\begin{figure}
\begin{center}
\vspace{1cm}
(a)\includegraphics[scale=0.25,angle=270]{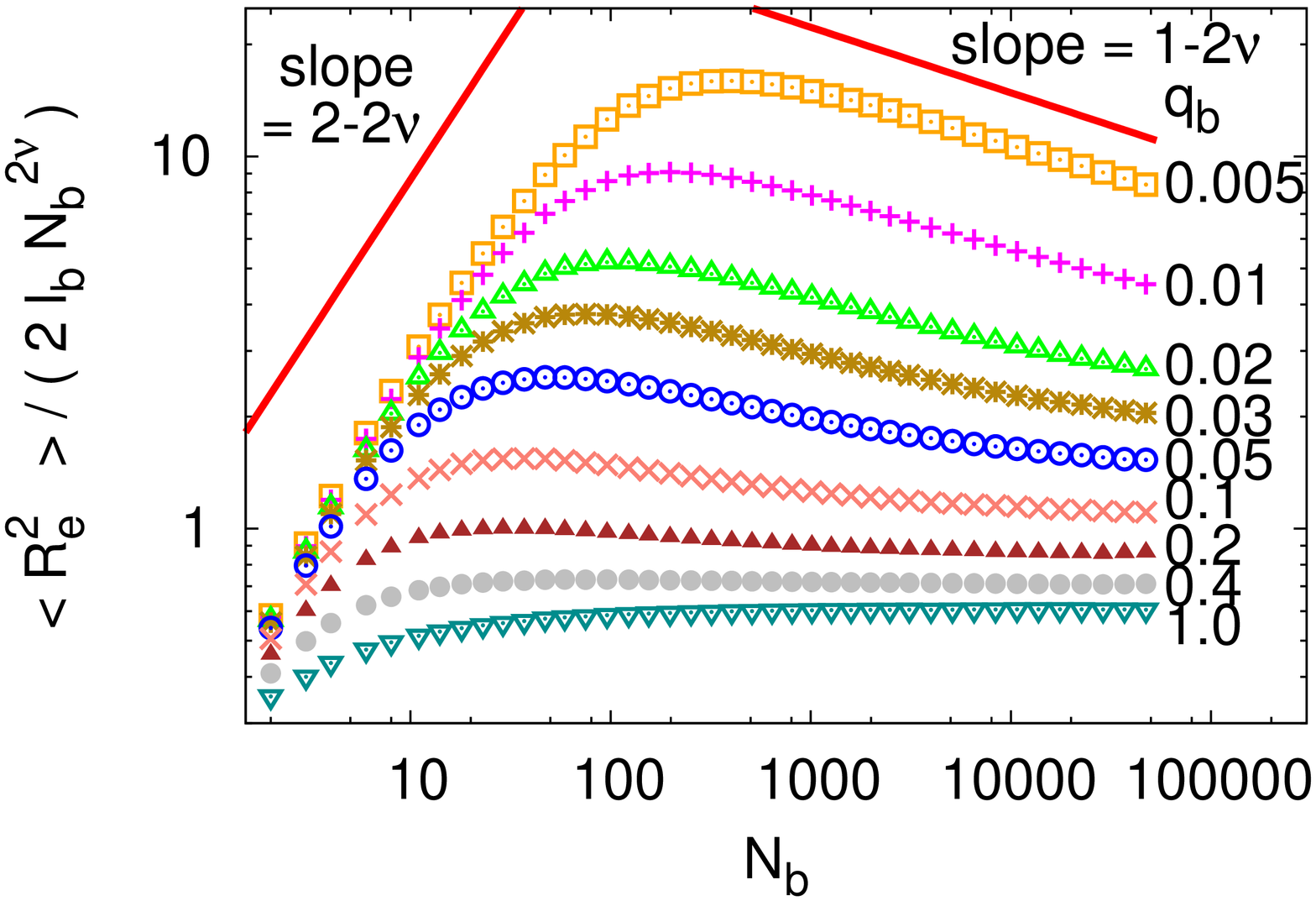}\\
(b)\includegraphics[scale=0.25,angle=270]{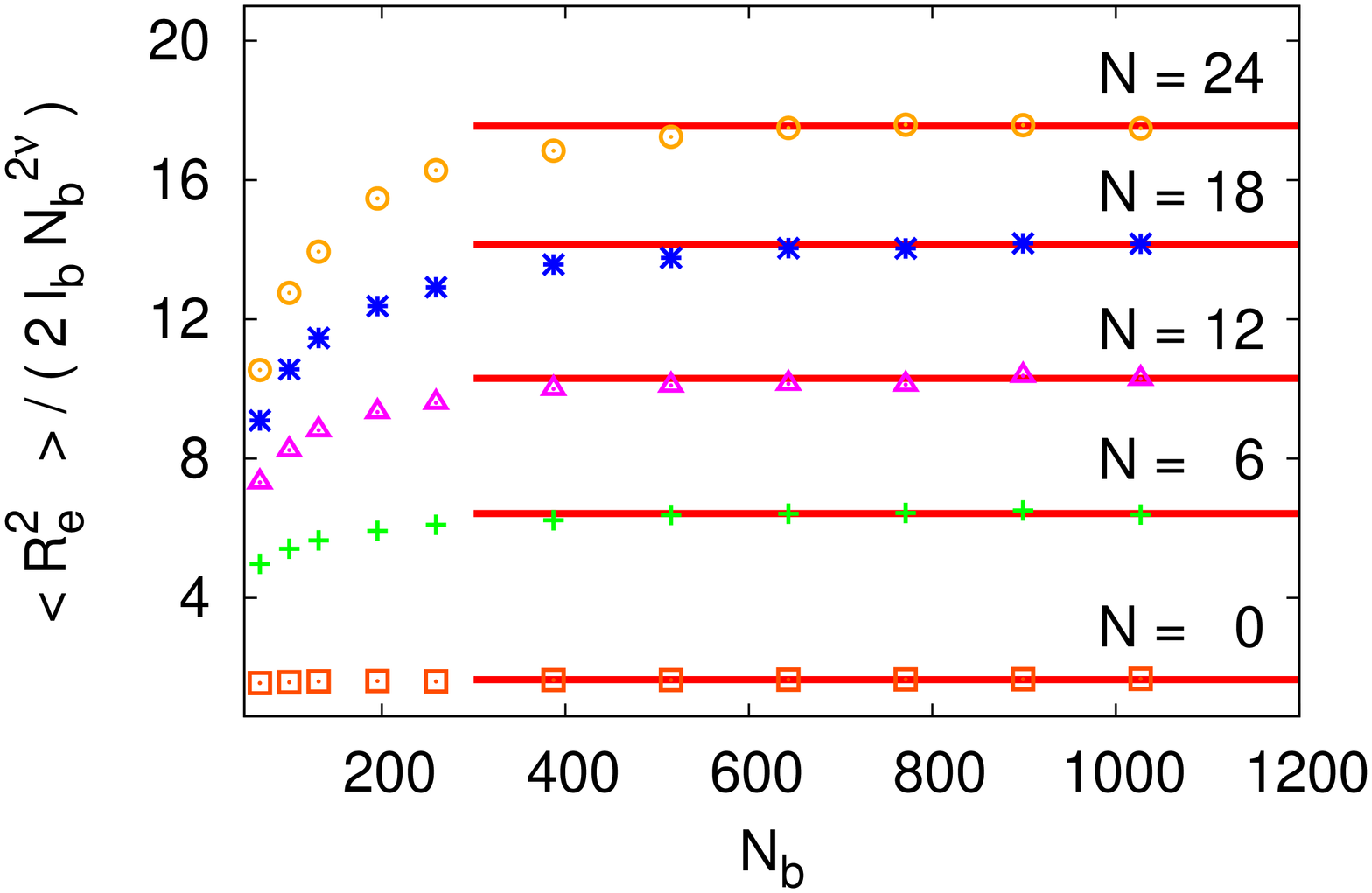}
\caption{Rescaled mean square end-to-end distance $\langle R^2_e
\rangle /(2 \ell_b N_b^{2 \nu})$ plotted against the chain length
$N_b$ for semiflexible chains with $\ell_b=1$ and variable $q_b$
(and hence $\ell_p)$ on a log-log plot (a), and analogous quantity
for the backbone of a bottle-brush polymer \cite{25}, using the
bond fluctuation model with excluded volume interactions only
which has $\ell_b=2.7$ lattice spacings,
and various side chain lengths $N$, on a linear-linear plot (b).
The slopes shown in (a) illustrate the exponents expected for rods
$(\langle R^2_e \rangle \propto N^2_b)$ or Gaussian chains
$(\langle R^2_e \rangle \propto N_b$), respectively.
Lengths are measured in units of the lattice spacing throughout.} \label{fig2}
\end{center}
\end{figure}

The chain linear dimensions tell a similar story
(fig.~\ref{fig2}): for flexible chains, $q_b=1.0$ and $q_b=0.4$, there
is a monotonic increase towards the plateau that yields the value
$\ell_{p,R} $ (eq.~(\ref{eq5})), fig.~\ref{fig2}a. However, the
resulting values for $\ell_{p,R}$ are small, unlike the case of
bottle-brushes (fig.~\ref{fig2}b), with one side chain
per backbone monomer, where with increasing side
chain length $N$ rather large values of $\ell_{p,R}$ can be
achieved \cite{25}. The strong initial rise seen for small $q_b$
in fig.~\ref{fig2}a, however, is indeed compelling evidence that
for small $q_b$ a rod-like behavior is found. But also for large
$q_b$ in fig.~\ref{fig2}a, and also in the case of the
bottle-brushes (fig.~\ref{fig2}b), 
one observes for small $N_b$ a strong increase of
$\langle R^2_e \rangle$ with $N_b$, but not as strong as it would
occur for rods.

\begin{figure}
\begin{center}
\vspace{1cm}
(a)\includegraphics[scale=0.25,angle=270]{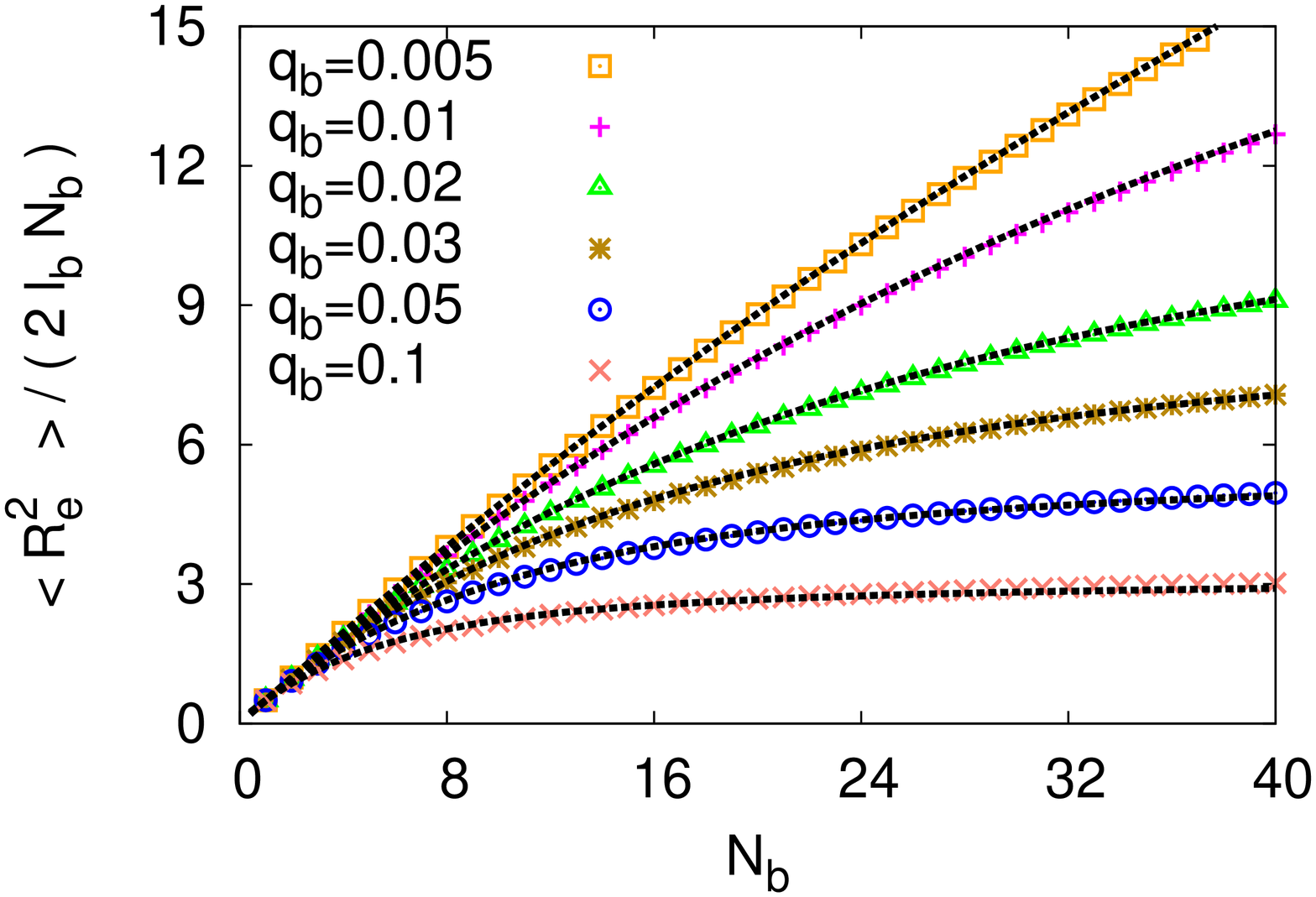}\\
(b)\includegraphics[scale=0.25,angle=270]{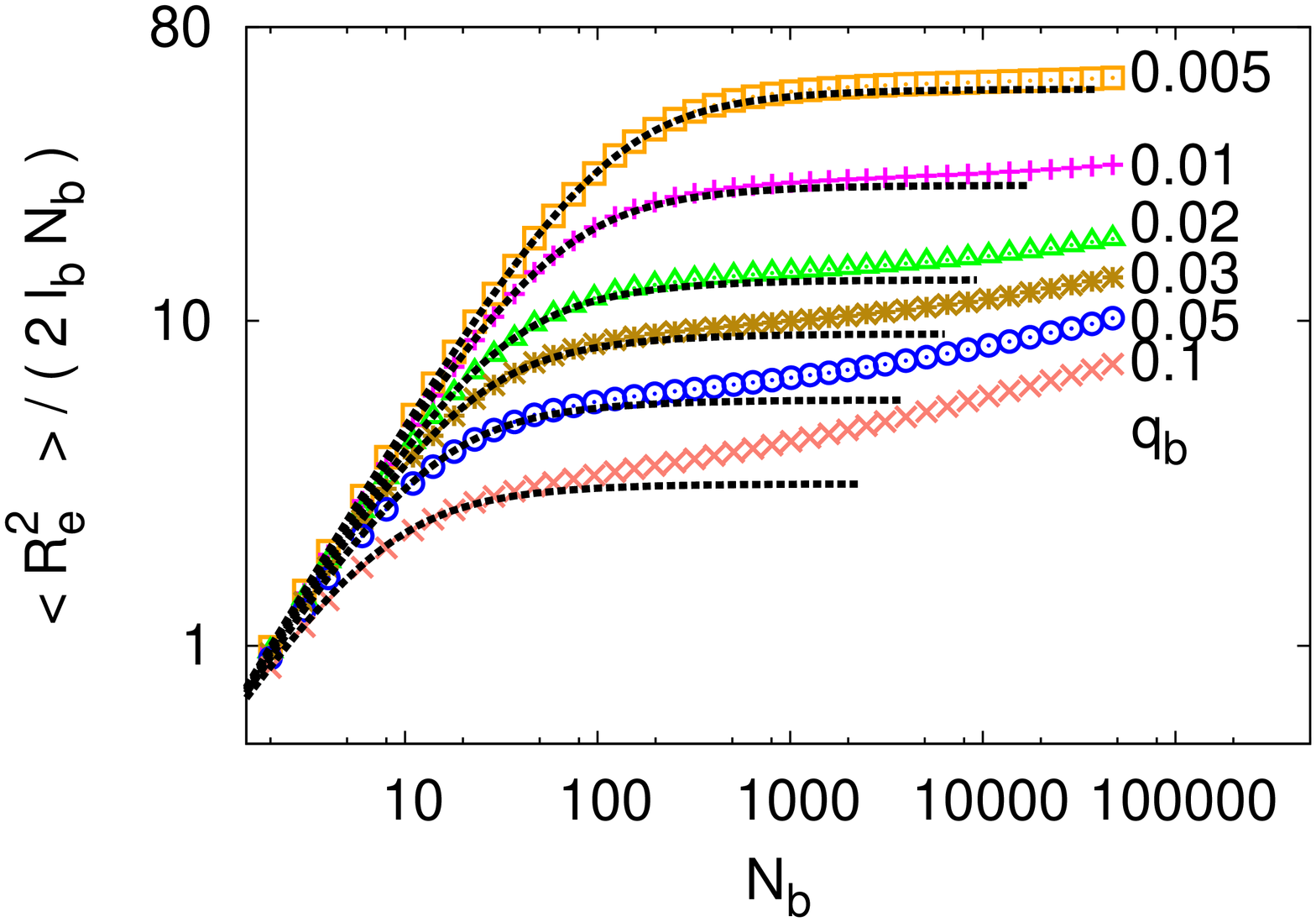}
\caption{Plot of $\langle R^2_e \rangle / 2 \ell_b N_b$ vs. $N_b$,
on linear-linear scales, for $N_b \leq40$ (a) and on log-log
scales, up to $N_b=50000$ (b), using only the data for $0.005 \leq
q_b \leq 0.1$. Dotted curves refer to the discrete chain model
of ~\cite{21}.}
\label{fig3}
\end{center}
\end{figure}

Fig.~\ref{fig3} now tests the applicability of the Kratky-Porod
formula, eq.~(\ref{eq3}), to these data; extracting $\ell_p$ from
the exponential fit of eq.~(\ref{eq2}) to the data in
fig.~\ref{fig1}c, there is no adjustable parameter whatsoever!
Here the data for $q_b \geq 0.2$ were omitted, since they clearly
do not show any trace of the Gaussian behavior implied by
eq.~(\ref{eq3}). For small $N_b$ and small $q_b$ the success of
eq.~(\ref{eq3}) indeed is remarkable (fig.~\ref{fig3}a), however,
for large $N_b$ we see that the data do not really settle down at
the ``plateau'' value implied by the Gaussian behavior, but rather
start to rise again, the crossover to the excluded volume behavior
sets in. Again, this crossover is spread out over about a decade
in $N_b$ (and for $q_b=0.005$ it starts only for $N_b > 50000!$).

\begin{figure}
\begin{center}
\vspace{1cm}
(a)\includegraphics[scale=0.25,angle=270]{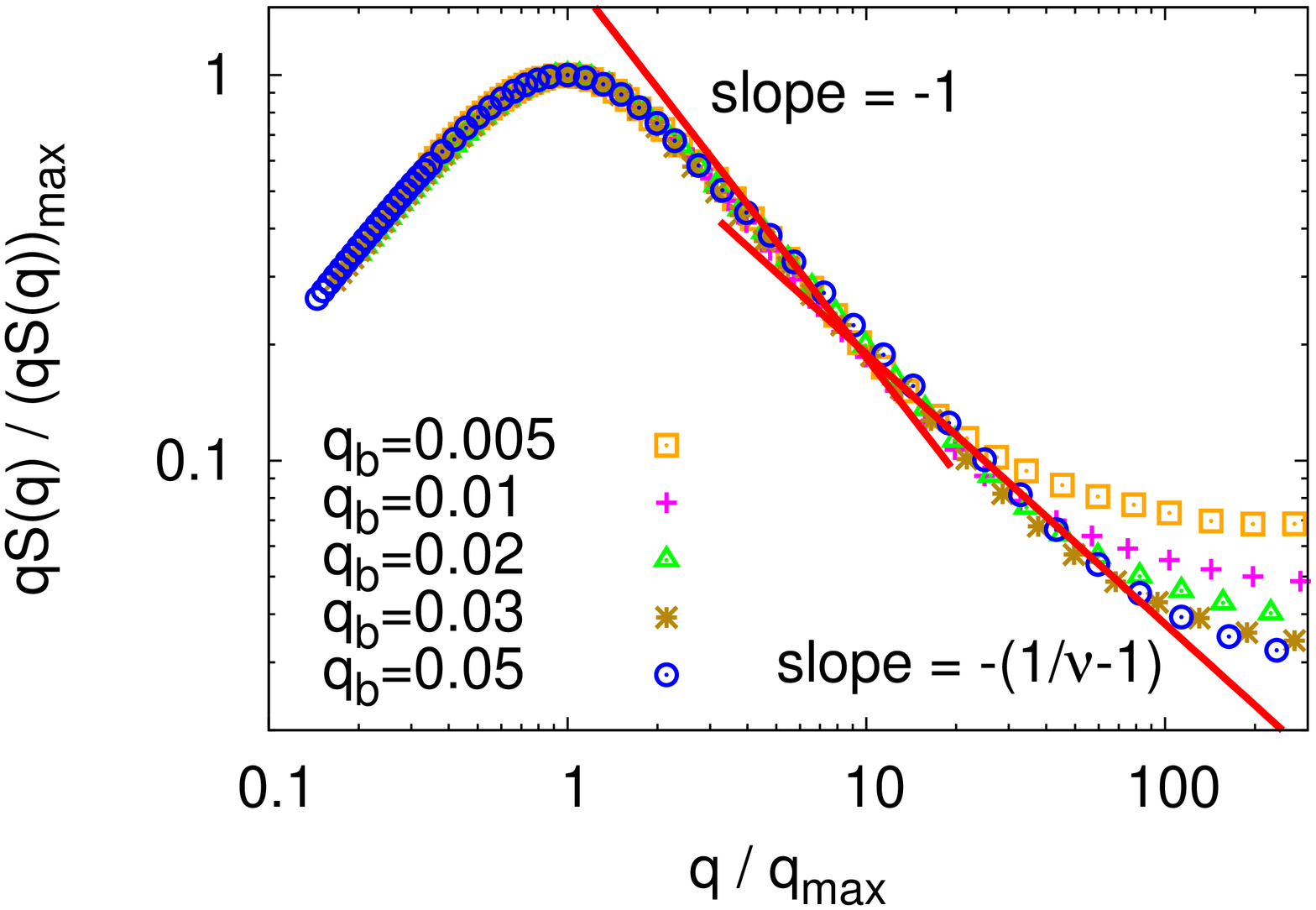}\\
(b)\includegraphics[scale=0.25,angle=270]{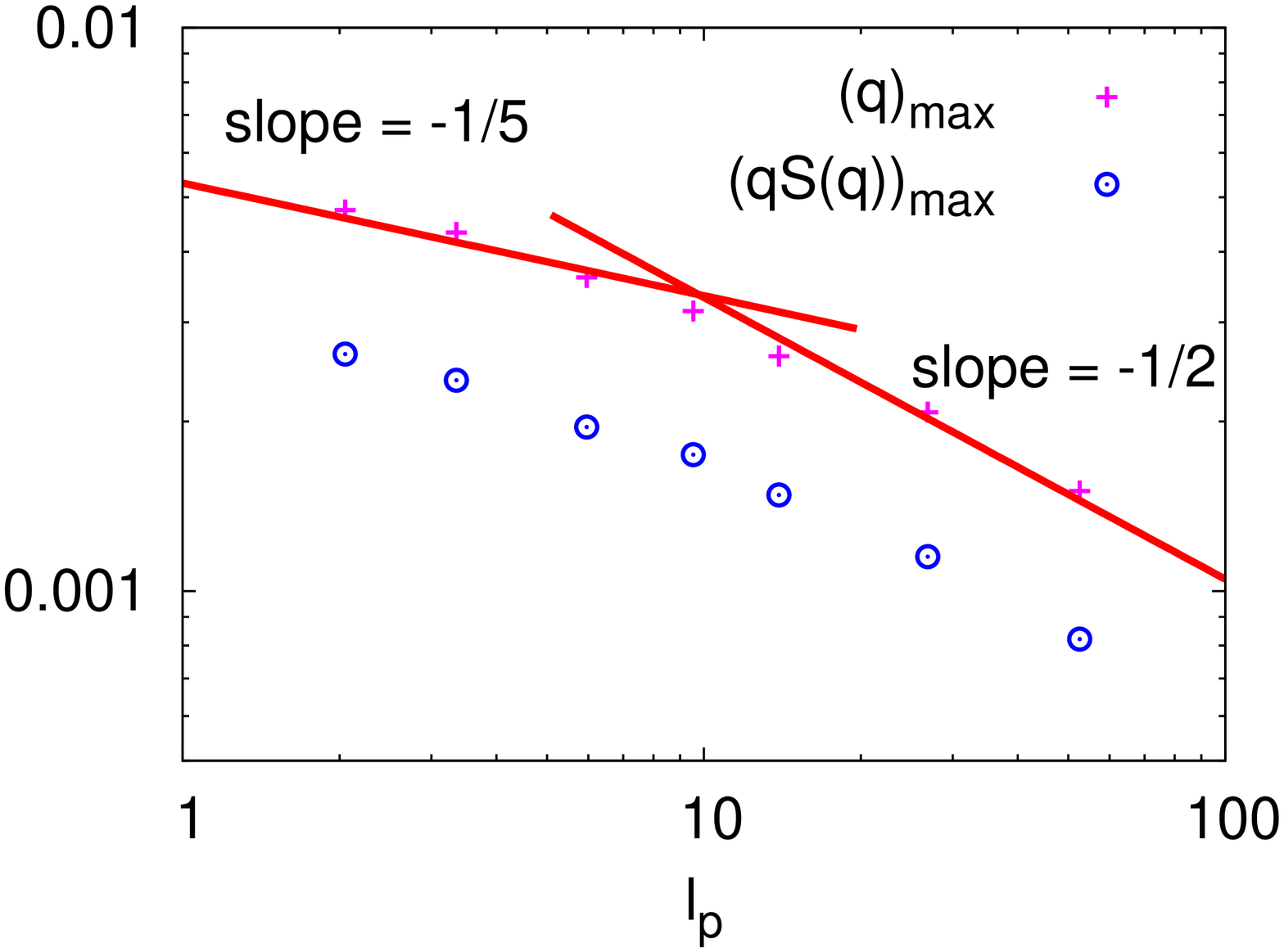}
\caption{(a) Rescaled Kratky-plot, log-log plot of $qS(q)/[q
S(q)]_{\rm \max}$ vs. $q/q_{\rm max}$, for several choices of
$q_b$ as indicated. Straight line indicates the effective exponent
that occurs in the regime of intermediate wavenumbers $q$. The
position $q_{\rm max}$ where $q S(q)$ exhibits its maximum $[q
S(q)]_{\rm max}$, and the height of this maximum is plotted in (b)
as function of $\ell_p$. Straight lines indicate the exponents
$q_{\rm max} \propto \ell_p^{-1/5}$ and $q_{\rm max} \propto
\ell_p^{-1/2}$ that one expects according to Netz and Andelman
\cite{28} in the excluded volume regime and Gaussian regime,
respectively. All data have been taken for $N_b=50000$.}
\label{fig4}
\end{center}
\end{figure}

In experiments, information on chain stiffness of polymers, in
principle, is accessible from an analysis of the static structure
factor $S(|\vec{q}|)$ of single chains where $\vec{q}$ is the
scattering vector. From a ``Kratky plot'', $q S(q)$ vs. $q$, one
finds a position where $S(q)$ is maximal, 
$q_{\rm max} \propto 1/R_{\rm gyr}$ at small $q$,
where $R_{\rm gyr}$ is the gyration radius of the polymer,
and a power law $q S(q) \propto q ^{-(1/\nu-1)}$ for chains
exhibiting excluded volume statistics at larger $q$. For very stiff
chains, one would expect that this power law crosses over to the
Gaussian behavior $q S(q) \propto q^{-1}$ at large $q$ and
ultimately a further crossover to the scattering function from
straight rods, $q S(q) = {\rm const.}$, should occur.
fig.~\ref{fig4}a shows that the latter regime is only found for
the stiffest chains that we could study ($q_b=0.005$), and the
onset of this so-called ``Holtzer plateau'' \cite{31} in the
Kratky plot is rather gradual: contrary to some suggestions
\cite{32,33}, estimation of the value $q^*$ where this onset
occurs is not suited as an accurate method to extract $\ell_p
(=1/q^*)$. Fig.~\ref{fig4}a also shows that the expected power
laws cannot be clearly identified either, because the crossover
between them is too gradual. Fig.~\ref{fig4}b reveals that also
the position $q_{\rm max}$ where $ q S(q)$ has its maximum does
not (yet) exhibit the expected power laws~\cite{28}: the
reason is that for small $q_b$ the chain length $N_b=50000$ still is
by far too short that the mean square radii $\langle R^2_e
\rangle$ (cf. fig.~\ref{fig2}a) and $\langle R^2_{\rm gyr}
\rangle$ can reach their asymptotic behavior. The smaller $q_b$
becomes (and the larger hence $\ell_p$ is) the more Gaussian
character of the chains do we expect for fixed $N_b=50000$, and
hence we see a gradual crossover from $q_{\rm max}
\propto\ell_p^{-1/5}$ to $q_{\rm max} \propto \ell_p^{-1/2}$ with
increasing $\ell_p$ on the log-log plot.

\begin{figure}
\begin{center}
\vspace{1cm}
(a)\includegraphics[scale=0.25,angle=270]{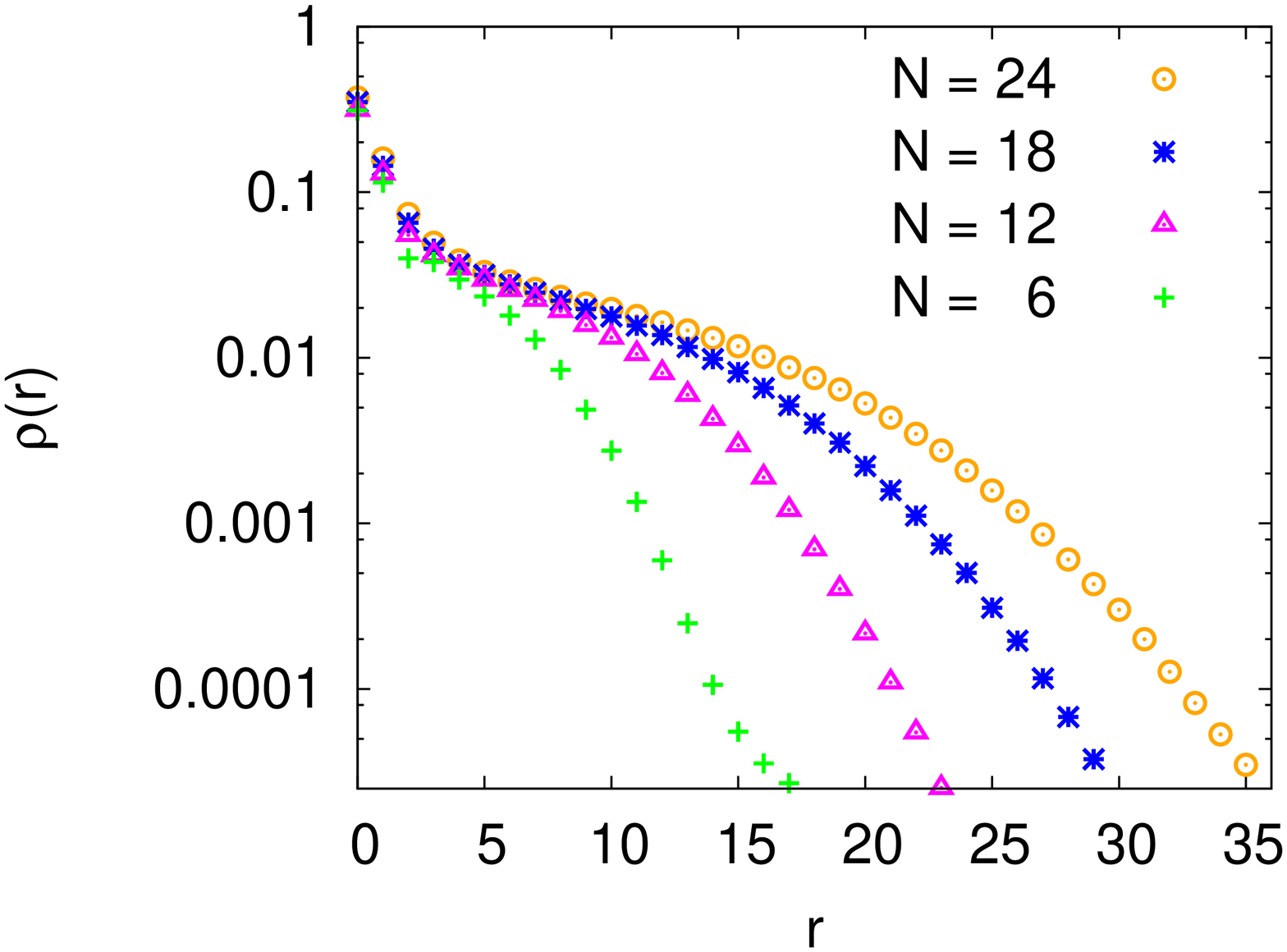}\\
(b)\includegraphics[scale=0.25,angle=270]{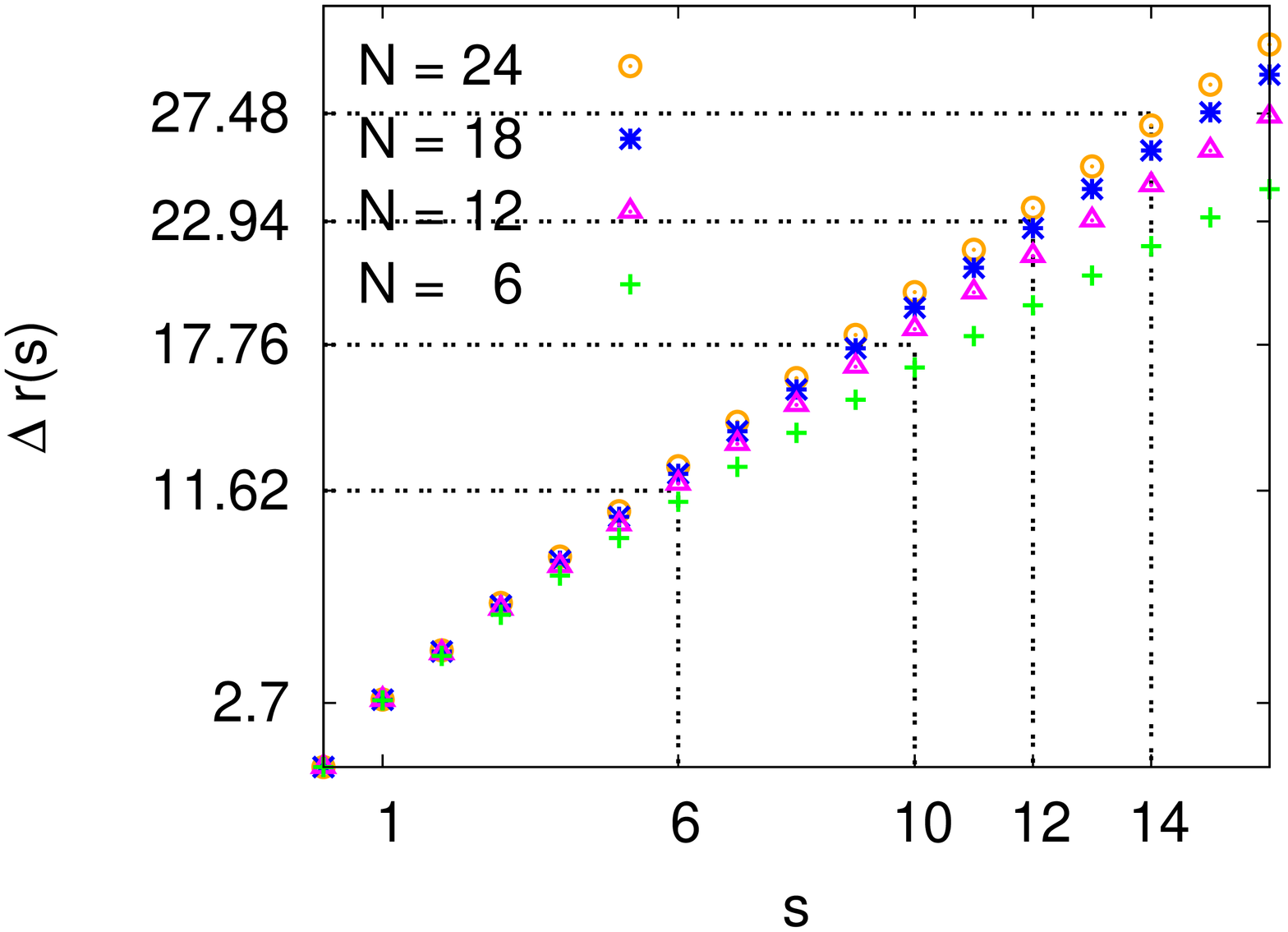}\\
(c)\includegraphics[scale=0.25,angle=270]{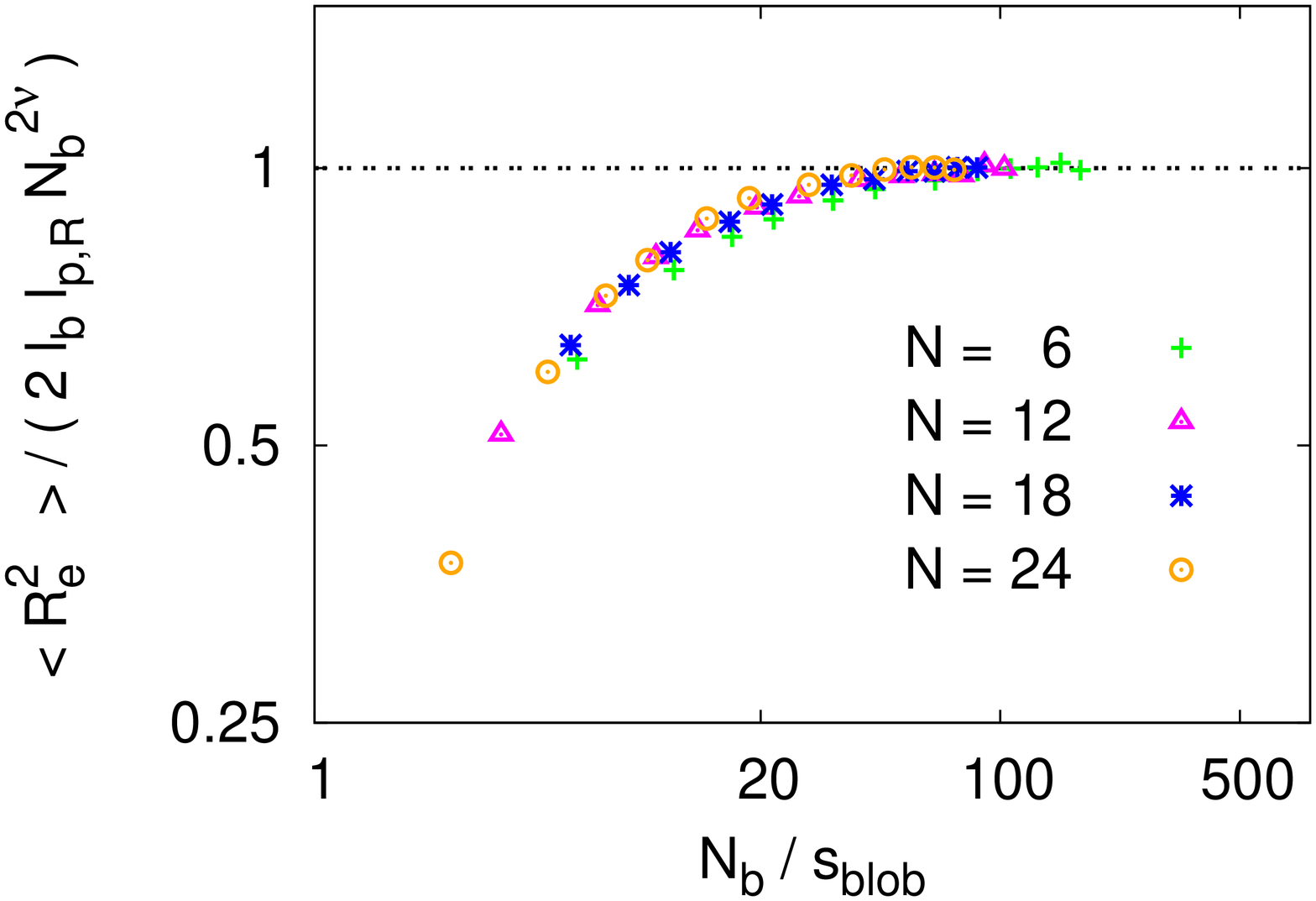}
\caption{(a) Radial monomer density distribution $\rho(r)$ in
planes locally perpendicular to the backbone of bottle-brush
polymers with backbone lengths $N_b=1027$ plotted vs. distance $r$
for side chain lengths $N=24$, $18$, $12$ and $6$. (b) Spatial distance
$\Delta r(s)$ between monomers a chemical distance $s$ apart plotted on a
double-logarithmic scale. The construction
$\Delta r(s)=2R_{cs}(N)$ is shown. (c) Rescaled mean square
end-to-end distance of bottle-brush backbones, $\langle
R^2_{e,b} \rangle / (2 \ell_b \ell_{p,R} N_b^{2 \nu})$ plotted
vs. the rescaled chain length $N_b/s_{\rm blob}$ as described in
the text.
Note that our length unit (lattice spacing) physically corresponds
to about $0.3$nm~\cite{35}} \label{fig5}
\end{center}
\end{figure}

Returning now to the behavior of bottle-brush polymers
(fig.~\ref{fig2}b), we emphasize that pronounced local stiffness
occurs there by a completely different mechanism, namely as a
consequence of the local thickening of the (coarse-grained)
cylindrical chains. This thickening with increasing side chain
length $N$ is evident from the radial density distribution in the
plane (locally \cite{34}) perpendicular to its backbone
(fig.~\ref{fig5}a). Normalizing $\rho(r)$ in fig.~\ref{fig5}a
such that $N=2 \pi \int\limits_0^\infty \rho(r) r dr $, the cross
-sectional radius $R_{\rm cs}(N)$ can be defined as $R^2_{\rm cs}= 2 \pi
\int \rho(r) r^3 dr$. We now view the bottle-brush polymer as a
sequence of blobs with diameter $2 R_{\rm cs} (N)$. Fig.~\ref{fig5}b shows the
construction. On the ordinate the values for $2R_{cs}(N)$ are indicated, and we
can identify how many backbone monomers $s_{\rm
blob} (N)$ occur per blob by requiring $\Delta r(s_{\rm
  blob}(N))=2R_{cs}(N)$. We find $s_{\rm blob}=6$, $10$, $12$ and $14$, 
for $N=6$, $12$, $18$ and $24$, respectively. If we rescale $N_b$ with
$s_{\rm blob}$, and $\langle R^2_{e,b} \rangle$ in fig.~\ref{fig2}b with
$2 \ell_b \ell_{p,R} N^{2\nu}_b$, we find that all data of
fig.~\ref{fig2}b fall on a universal curve (fig.~\ref{fig5}b)!.
The scaling implies that for bottle-brush polymers the large
values of the effective persistence lengths $\ell_{p,R}$ are
entirely due to the side-chain induced thickening of these
flexible cylindrical brushes, at least for the range of side chain
lengths accessible in the simulations. But this range nicely
coincides \cite{35} with the range accessible in experiments
\cite{9,11,13,14,15,32,33,36}.

Our findings that the persistence length $\ell_p$ of bottle-brush
polymers is of the same order as the cross-sectional radius
$R_{\rm cs}$ agrees with the result of Birshtein et al. \cite{37} but
disagrees with later scaling theories \cite{10,38}. In fact, if
$\ell_p \propto R_{\rm cs}$ the second virial coefficient $\upsilon_2$
scales as \cite{28} $\upsilon_2 \propto \ell_p^2 R_{\rm cs} \propto
\ell^3_p$, and the Flory theory would yield an end-to-end distance
\cite{28} $R_e \propto(\upsilon_2/\ell_p)^{1/5} \ell_b^{3/5}
N^{3/5}_b=\ell^{2/5}_p \ell^{3/5}_b N^{3/5}_b$. A Gaussian
behavior, $R^2_e=\ell_b \ell_p N_b$, is only expected to
occur for $\ell_p/\ell_b < N_b <N_b^*$. However, equating the
above two expressions for $R_e$ yields $N_b^*\propto
\ell_p/\ell_b$ as well, i.e., in agreement with our findings, the
theory of Netz and Andelman \cite{28} also implies that a
preasymptotic Gaussian regime does not occur for
bottle-brushes. The scaling theories on the contrary imply that $R_{\rm cs} \propto
N^{3/4}$ and $\ell_p \propto N^{15/8}$, and hence clearly
would predict the existence of a Gaussian regime. Furthermore, even for much
larger side chains the behavior $R_{\rm cs} \propto N^{ 3/4}$ could
not be verified \cite{39}. Although we clearly cannot exclude that
the scaling theory might become valid for side chain lengths of
the order of $N \approx 10^3$ or larger, this clearly is
completely irrelevant for both experiments
\cite{9,11,13,14,15,32,33,36} and simulations \cite{25,34,35,39}.
In fact, self-consistent field calculations~\cite{43} have shown
that side chain interactions do have an important effect
onto the apparent persistence length only if
$N \ge 1000$.

\section{Conclusions}

In summary, we have verified two distinct scenarios for local
stiffness of polymer chains: if chain rigidity is caused by an
energy term that suppresses chain bending to a large extent,
without increase of the cross-sectional radius of the chains, one
finds a regime of wormlike chain behavior, described by the
Kratky-Porod model \{e.g., eq.~(\ref{eq3})\} for chains up to a
chain length $N^*_b$ (where $N^*_b$ can be estimated by a Flory
argument, cf. eq.~(\ref{eq6}) \cite{28}), where a gradual
crossover to excluded-volume-dominated behavior
\{eq.~(\ref{eq5})\} occurs. However, if chain rigidity is caused
by an increase of local chain thickness (as occurs for
bottle-brush polymers), no preasymptotic Gaussian regime
compatible with the Kratky-Porod model occurs. However, another
preasymptotic regime occurs, for $2 < N_b/s_{\rm blob} < 20$,
where $\langle R^2_{e,b} \rangle$ gradually crosses
over from the rod-like behavior $\langle R^2_{e,b} \rangle \propto
N^2_b$, to the excluded volume behavior, $\langle R^2_{e,b}
\rangle \propto N^{2 \nu}_b$ (fig.~\ref{fig5}b). 
In a regime where the contour length exceeds the persistence 
length only a few times, the Kratky-Porod model can be fitted
to experimental data, excluded volume not
yet being very important. 
However, other experiments may fall in a regime where 
excluded volume effects are important~\cite{44,45,46}.
In ~\cite{45,46} the importance of excluded volume was inferred from
the concentration dependence in semi-diilute solution.
Clearly, it is important to carefully examine to which regime 
experimental data belong.

\underline{Acknowledgement}: This work has been supported by the
Deutsche Forschungsgemeinschaft (DFG) under grant No SFB625/A3 and
by the European Science Foundation (ESF) under the STIPOMAT
program. We thank the J\"ulich Supercomputing Centre for computer
time at the NIC Juropa under the project No HMZ03
and the SOFTCOMP clusters, and are grateful
to S. Rathgeber, M. Schmidt, and A. N. Semenov for stimulating
discussions.


\begin{thebibliography}{99}
\bibitem{1} 
  {GROSBERG A. Yu. and  KHOKHLOV A. R.}, 
  {Statistical Physics of Macromolecules}, 
  {AIP Press, New York}
  {(1994)}.

\bibitem{2}
  {DES CLOIZEAUX J. and JANNINK G.}, 
  {Polymers in Solution: Their Modeling and Structure.}
  {Clarendon Press, Oxford}
  {(1990)}.

\bibitem{3}
  {RUBINSTEIN M. and  COLBY R. H.},
  {Polymer Physics}
  {Oxford Univ. Press, Oxford}
  {(2003)}.

\bibitem{4}
  {DE GENNES P. G.},
  {Scaling Concepts in Polymer Physics} ,
  {Cornell University Press, Ithaca, N.Y.}
  {(1979)}.

\bibitem{5}
  {SCH\"AFER L.},
  {Excluded Volume Effects in Polymer Solutions as Explained
  by the Renormalization Group},
  {Springer, Berlin}
  {(1999)}.

\bibitem{6}
  {BUSTAMANTE C., MARKO J. F., SIGGIA E. D. and SMITH, S.},
  {Science} {265} {(1994)} {1599}.

\bibitem{7}
  {K\"AS J., STREY H., TANG J. X., FINGER D., EZZELL R., SACKMANN E.
 and JANMEY P. A.},
  {Biophys J.} {70} {(1996)} {609}.

\bibitem{8}
  {OBER C. K.},
  {Science} {288} {(2000)} {448}.

\bibitem{9}
  {ZHANG M. and M\"ULLER A. H. E.},
 {J. Polym. Sci.: Part A: Polm. Chem.} {43} {(2005)} {3461}.

\bibitem{10}
 {SUBBOTIN A. V. and SEMENOV A. N.},
 {Polymer Science, Ser. A} {49} {(2007)} {1328}.

\bibitem{11}
  {SHEIKO S. S., SUMERLIN B. S. and MATYJASZEWSKI K.},
  {Progr. Polym. Sci.} {33} {(2008)} {759}.

\bibitem{12}
  {POTEMKIN I. I. and PALYULIN V. V.},
 {Polymer Science, Ser. A.} {51} {(2009)} {123}.

\bibitem{13}
  {WINTERMANTEL M., FISCHER K., GERLE M., RIES R., SCHMIDT M.,
 KAJIWARA K., URAKAWA H. and  WATAOKA I.},
  {Angew. Chem., Int. Ed.} {34} {(1995)} {1472}.

\bibitem{14}
  {STEPHAN T., MUTH S. and SCHMIDT M.},
 {Macromolecules} {35} {(2002)} {9857}.

\bibitem{15}
  {LI C.,  GUNARI N., FISCHER K., JANSHOFF A. and SCHMIDT M.},
 {Angew. Chem., Int. Ed.} {43} {(2004)} {1101}.

\bibitem{16}
 {IOZZO R. V. (ed.)},
 {Proteoglycans: Structure, Biology, and Molecular Interactions} ,
 {Marcel Dekker, New York} 
 {(2000)}.

\bibitem{17}
  {KLEIN J.},
 {Science} {323} {(2009)} {47}.

\bibitem{18}
 {KRATKY O. and POROD G.},
 {J. Colloid Sci.} {4} {(1949)} {35}.

\bibitem{19}
  {BAWENDI M. G. and FREED K. F.},
 {J. Chem. Phys.} {83} {(1985)} {2491}.

\bibitem{20}
 {LAGOWSKI J. B., NOOLANDI J. and NICKEL B.},
{J. Chem. Phys.} {95} {(1991)} {1266}.

\bibitem{21} 
  {WINKLER R. G.,  REINEKER, P. and HARNAU L.},
 {J. Chem. Phys.} {101} {(1994)} {8119}.

\bibitem{22}
  {STEINHAUSER M. O., SCHNEIDER J. and BLUMEN A.},
 {J. Chem. Phys.} {130} {(2009)} {164902}

\bibitem{23}
  {SCH\"AFER L., OSTENDORF A. and HAGER J.},
 {J. Phys. A: Math. Gen.} {32} {(1999)} {7875}.

\bibitem{24}
 {LE GUILLOU and ZINN-JUSTIN J.},
 {Phys. Rev. B} {21} {(1980)} {3976}.

\bibitem{25}
 {HSU H.-P., PAUL W., BINDER K.}, 
 {Macromolecules} {43} {(2010)} {3094}.

\bibitem{26}
{An analoguous hypothesis was suggested for chains at the Theta
point, where $\beta=3/2$} by
  {SHIRVANYANTS D., PANYUKOV S., LIAO Q., RUBINSTEIN M.},
  {Macromolecules} {41} {(2008)} {1475}.

\bibitem{27}
  {FLORY, P. J.},
  {Principles of Polymer Chemistry},
  {Cornell Univ.  Press, Ithaca, N.Y.} 
  {(1953)}.

\bibitem{28}
  {NETZ R. R. and ANDELMANN D.},
  {Phys. Repts} {380} {(2003)} {1}.

\bibitem{29a} 
  {SCHAEFER D. W., JOANNY J. F. and PINCUS P.},
  {Macromolecuels} {13} {(1980)} {1280}.

\bibitem{30a}
  {BIRSHTEIN T. M.},
  {Polymer Sci. (USSR)} {A24} {(1982)} {2416}.

\bibitem{31a}
  {BIRSHTEIN T. M. and ZHULINA E. B.},
  {Polymer} {25} {(1984)} {1453}.
 
\bibitem{29}
  {GRASSBERGER P.}
  {Phys. Rev. E} {56} {(1997)} {3682}.

\bibitem{30}
  {BASTOLLA U. and GRASSBERGER P.}
  {J. Stat. Phys.} {89} {(1997)} {1061}.

\bibitem{31}
 {HOLZER A.}
 {J. Polym. Sci.} {17} {(1955)} {432}.

  \bibitem{32}
  {LECOMMANDOUX S., CHII{\'E}COT F., BARSALI R., SCHAPPACHER M., 
  DEFFIEUX A., BR{\^U}LET A. and COTTON J. P.}, 
  {Macromolecules} {35} {(2002)} {8878}.

  \bibitem{33}
  {ZHANG B., GR\"OHN F., PEDERSEN J. S., FISCHER K. and SCHMIDT M.},
  {Macromolecules} {39} {(2006)} {8440}.

 \bibitem{34}
  {HSU H.-P., BINDER K. and  PAUL W.},
  {Phys. Rev. Lett.} {103} {(2009)} {198301}.

 \bibitem{35}
  {HSU H.-P., PAUL W., RATHGEBER S. and BINDER K.},
  {Macromolecules} {43} {(2010)} {1592}.

 \bibitem{36}
  {RATHGEBER S., PAKULA T., MATHYJASZEWSKI K. and BEERS K.L.}
  {J. Chem. Phys.} {122} {(2005)} {124904}.

 \bibitem{37}
  {BIRSHTEIN T. M., BORISOV O. V., ZHULINA E. B.
  and  YURASOV T. A.},
  {Polym. Sci. USSR} {29} {(1987)} {1293}.

\bibitem{38}
  {FREDRICKSON G. H.},
  {Macromolecules} {26} {(1993)} {7214}.

\bibitem{39}
  {HSU H.-P., PAUL W. and BINDER K.},
  {Macromol. Theory Simul.} {16} {(2007)} {660}.

\bibitem{43} 
  {FEUZ L., LEERMAKERS F. A. M., TEXTOR M. and BORISOV O. V.},
  {Macromolecules} {38} {(2005)} {8891}.

\bibitem{44}
  {CHENG G., MELNICHENKO Y. B., WIGNALL G. D., HUA F.,
        HONG K. and MAYS J. W.},
  {Macromolecules} {41} {(2008)} {9831}.

\bibitem{45}
  {BOLISETTY S., ROSENFELDT S., ROCHETTE C. N., HARNAU L.,
        LINDNER P., XU Y., M\"ULLER A. H. E. and BALLAUFF M.},
  {Colloid Polym. Sci.} {287} {(2009)} {129}.

\bibitem{46} 
  {Bolisetty S., Airaud C., Xu Y., M\"uller A. H. E., Harnau L.,
Rosenfeldt S., Lindner P. and Ballauff M.},
  {Phy. Rev. E} {75} {(2007)} {040803(R)}.

  \end{thebibliography}
\end{document}